\documentclass[modern]{aastex61}

\submitjournal{ApJ}

\shorttitle{Kpc-scale jets}
\shortauthors{Kusunose \& Takahara}
\begin{document}

\title{A Photo-Hadronic Model of the Large Scale Jets of 3C 273 and PKS 1136--135}

\correspondingauthor{Masaaki Kusunose}
\email{kusunose@kwansei.ac.jp, takahara.fumio@wine.plala.or.jp}

\author[0000-0002-2465-7571]{Masaaki Kusunose}
\affiliation{Department of Physics, Kwansei Gakuin University\\
  Gakuen  2-1, Sanda \\
  Hyogo, 669-1337, Japan}
%\nocollaboration

\author{Fumio Takahara}
\affiliation{Department of Earth and Space Science, Graduate School of Science,
  Osaka University \\
  Machikaneyama 1-1, Toyonaka, Osaka 560-0043, Japan}
%\nocollaboration

%% Note that the \and command from previous versions of AASTeX is now
%% depreciated in this version as it is no longer necessary. AASTeX 
%% automatically takes care of all commas and "and"s between authors names.

%% AASTeX 6.1 has the new \collaboration and \nocollaboration commands to
%% provide the collaboration status of a group of authors. These commands 
%% can be used either before or after the list of corresponding authors. The
%% argument for \collaboration is the collaboration identifier. Authors are
%% encouraged to surround collaboration identifiers with ()s. The 
%% \nocollaboration command takes no argument and exists to indicate that
%% the nearby authors are not part of surrounding collaborations.

%% Mark off the abstract in the ``abstract'' environment. 
\begin{abstract}
  X-ray bright knots of kpc-scale jets of several radio loud quasars have been 
  an actively discussed issue. 
  Among various models to explain observations, synchrotron radiation 
  from the electron population different from radio to IR emitting electrons
  is promising.
  However, the origin of this electron population has been debated.
  Recently, we proposed that this electron population is produced by proton-photon
  collisions (mainly, Bethe-Heitler process), and we applied this model to PKS 0637--752.
  We found that this model works 
  if the proton power is by an order of magnitude larger than the Eddington power. 
  In this paper we apply this model to the X-ray emission in the knots of 
  3C 273 and PKS 1136--135.
  The target photons for electron-positron pair production are supplied 
  by synchrotron radiation at radio-IR  by primary electrons and 
  by the active galactic nucleus (AGN) core as well as cosmic microwave background (CMB) 
  radiation.
  The effects of the AGN photons are included for the first time 
  in the hadronic model.
  Though the observed X-ray flux is obtained 
  with the contribution of the AGN photons, 
  the required proton power turns out to be highly super-Eddington.
  However, we find that our model works for a nearly Eddington proton power,
  if the photon density of the AGN is enhanced.
  This can occur if the AGN photons are more beamed toward the X-ray knots than toward 
  our line of sight and the AGN photon frequency is shifted by the Doppler effect.
\end{abstract}

%% Keywords should appear after the \end{abstract} command. 
%% See the online documentation for the full list of available subject
%% keywords and the rultoward our line of sightes for their use.

%% From the front matter, we move on to the body of the paper.
%% Sections are demarcated by \section and \subsection, respectively.
%% Observe the use of the LaTeX \label
%% command after the \subsection to give a symbolic KEY to the
%% subsection for cross-referencing in a \ref command.
%% You can use LaTeX's \ref and \label commands to keep track of
%% cross-references to sections, equations, tables, and figures.
%% That way, if you change the order of any elements, LaTeX will
%% automatically renumber them.

%% We recommend that authors also use the natbib \citep
%% and \citet commands to identify citations.  The citations are
%% tied to the reference list via symbolic KEYs. The KEY corresponds
%% to the KEY in the \bibitem in the reference list below. 

\section{Introduction} \label{sec:intro}

Large scale jets of active galactic nuclei (AGNs) have been observed at various 
wavelengths.
The kpc-scale jets of some AGNs  are known to be bright in X-rays 
as observed with the \textit{Chandra X-Ray Observatory} 
\citep{schwartz2000,chartes2000}.
While the X-rays from FRI sources are usually explained by the extrapolation of 
the synchrotron spectrum at radio-IR emission,
the X-ray flux of some quasars is larger than the flux extrapolated from the radio-IR 
radiation.
Then various X-ray emission mechanisms for the kpc-scale jets have been proposed.
Synchrotron-self Compton scattering by nonthermal electrons, 
which is often accepted as X-ray emission model for AGN jets, is not favored,
because the required magnetic field is much weaker than the equipartition value 
\citep{chartes2000,ks2005}.
Inverse Compton (IC) scattering of cosmic microwave background (CMB) radiation 
in relativistically moving emission region beamed to the observer (IC/CMB model) 
was also considered \citep{tav2000,cel2001}.
However, the predicted flux by IC/CMB model exceeds the upper limits of 
the $\gamma$-ray flux of kpc jets in  PKS 0637--752 and 3C 273 observed by \textit{Fermi} 
\citep{mg2014,meyer2015}.
Additional $\gamma$-ray upper limits were recently presented for PKS 1136--135, 
PKS 1229--021, PKS 1354+195, and PKS 2209+080 \citep{bm2017}.
Proton synchrotron model was also proposed \citep{aha2002,bg2016},
which requires a rather large value of magnetic fields, e.g., 10 mG.
\footnote{\citet{aha2002} first examined the various processes involving ultra high energy protons.
He disfavored photo-hadronic processes because required proton power is too large. 
Instead he proposed proton synchrotron as reasonable 
but he assumed that protons escape the emission region diffusively.
Considering a large proton pressure, protons will be adiabatically cooled. 
Thus, energy requirement is not relieved.}
The most promising model is synchrotron radiation by two different 
populations of electrons/positrons 
\citep{uchi2006},
though different acceleration mechanisms for different electron/positron populations 
are to be considered.
\citet{liu2015} proposed shock regions in the knots of 3C 273:
Different electron populations are formed in the upstream and the downstream,
which correspond to radio-IR and X-ray emissions, respectively.
\citet{liu2017} recently proposed shear acceleration as a higher energy electron
production mechanism. 
Though these are interesting possibilities and feasible energetically,
detailed comparison between model prediction and observations is awaited.

Recently, to account for the second population of electrons,
we proposed an alternative model in which  high energy electrons/positrons 
produced by Bethe-Heitler process
and applied this model to PKS 0637--752 \citep{kt2017}.
Protons obeying a power-law spectrum produce electrons/positrons 
(hereafter we call shortly electrons)
in collisions with radio-IR photons emitted by synchrotron radiation in a knot.
Then produced electrons emit X-rays and $\gamma$-rays by synchrotron radiation.
We showed that X-ray emission is explained by this model with a size $\sim$ a few kpc
and a magnetic field $\sim 0.1$ mG.
However, the proton power an order of magnitude larger 
than the Eddington limit is required
because of the low efficiency of electron production in the radiation field given by
the radio-IR spectrum by primary electrons.
It is to be noted that though the conversion efficiency of photo-pion production 
processes is larger, 
required proton energy is much larger and the produced electrons emit synchrotron
photons at much higher energies than X-ray,
which makes photo-pion processes relatively ineffective.

There are more AGNs with kpc-scale jets bright in X-rays
and they  can be used to examine our model.
3C 273 is one of these sources,
a well known quasar at redshift $z=0.158$ \citep[see][for review]{uchi2006}.
Compared to PKS 0637--752, the luminosity of 3C 273 jet is lower and
the slope between optical and X-rays is softer.
However, the slope in the $\log \nu$-$\log \nu F_\nu$ representation
at $\sim 10^{15}$ Hz is still positive.
Because of the $\gamma$-ray upper limits set by \textit{Fermi}  \citep{meyer2015} and 
the spectral shape of IR-optical through X-rays as well as the optical polarization 
\citep{uchi2006},
the model with two electron populations is favored for 3C 273.
Regarding the size of the emission region of kpc-scale knots,
\citet{mar2017} recently analyzed the detailed structure of the jet in 3C 273, 
and the knots were resolved to have extended features with the transverse sizes of $\sim 0.5$ kpc 
and lengths $\gtrsim 1$ kpc in X-rays and far-ultraviolet.
This small size of the X-ray emission region is notable 
and it is worth studying the X-ray emission model by taking into account this small size,
though this may be regarded as tentative because the results 
depend on a sophisticated deconvolution technique.
The required proton power decreases with the decreasing size, 
when the target photon is provided internally.
Another notable feature of 3C273 jet is different brightness distribution between 
X-rays and radio. 
While radio emission is brighter at the outer edge of the jet, 
X-rays are stronger at inner portion. 
This may suggest photons from the AGN core play some role 
in shaping the X-ray emission from the large scale jet.
If the energy density of these photons in the knots dominates others, 
IC scattering by relativistic electrons and photo-hadronic processes 
are enhanced, which may somewhat relieve the energy problem.
This effect is worth studying, because this has not been considered in previous work.
Furthermore, photons from the AGN core are strongly beamed and the beamed 
direction is not necessarily peaked to the our line of sight. If the beaming 
is peaked toward the knots rather than our line of sight, the effects are much stronger.

Another kpc-scale jet is found in PKS 1136--135, 
a steep-spectrum quasar located at $z=0.556$.
The X-ray emitting knots of PKS 1136--135 are at projected distances of 30 - 60 kpc
from the core \citep{cara2013}.
\citet{sam2006} and \citet{tav2006} supported the IC/CMB model for the X-ray emission
from the knots.
On the other hand, \citet{uchi2007} extensively studied the emission mechanisms 
in this source and  interpreted the X-ray emission as synchrotron radiation 
from a second nonthermal electron population.
Later \citet{cara2013} found that several kpc-scale knots of PKS 1136--135
are highly polarized at optical.
Combining this high polarization and the spectral shape,
they concluded that IC/CMB models are not favored to explain the X-ray emission.

In this paper, we apply our model to the kpc-scale jets of 3C 273 and PKS 1136--135.
In particular, we study the effect of AGN photons on the electron/positron production 
and consider its effect on the size of the emission region.
In Section \ref{sec:model} we describe our model,
and in Section \ref{sec:results} numerical results are shown.
In Section \ref{sec:conclusion} summary and discussion are presented.
Throughout this paper we assume $\Omega_m = 0.27$, $\Omega_\Lambda = 0.73$,
$\Omega_r=0$, and $H_0 = 71$ km s$^{-1}$ Mpc$^{-1}$ for cosmological parameters.

%%%%%%%%%%%%%%%%%%%%%%%%%%%%%%%%%%%%%%%%%%%%%
\section{Assumptions and Formalism} \label{sec:model}

The model in this paper is the same as used in \citet{kt2017}
except that additional soft photon sources such as the AGN and CMB are considered.
We also include proton synchrotron radiation
as well as electron/positron production by photo-pion processes.
However, for simplicity we do not include $\gamma$-ray production via $\pi^0$ decay
and subsequent cascade processes. 
We assume a uniform sphere with radius $R$ for a knot with magnetic fields $B$.
The relativistic bulk motion of the  knot is not considered,
because there is no clear evidence for relativistic motion of kpc-scale jets in
3C 273 and PKS 1136--135 \citep[e.g.,][]{cara2013},
noting that only mildly relativistic beaming is enough to explain the asymmetry
between the jet and the counter-jet.
Radiation in radio through IR is emitted by nonthermal electrons in the knot.
These electrons are the primary particles different from the secondary electrons produced 
by Bethe-Heitler process.
The spectrum of these electrons is obtained by calculating the steady state solution
of the kinetic equation of electrons with injection, radiative cooling, and escape.
The kinetic equation for the primary electrons is given by
\begin{equation}
  \frac{d n_e (\gamma_e)}{d t}=q_\mathrm{inj}(\gamma_e) 
  -\frac{n_e (\gamma_e)}{t_\mathrm{esc}}
  -\frac{d~}{d\gamma_e} [\dot{\gamma}_e n_e (\gamma_e)] ,
  \label{eq:e-kinetic-hadronic}
\end{equation}
where $\gamma_e$ is the electron Lorentz factor
and $n_e(\gamma_e)$ is the electron number density per unit interval of $\gamma_e$.
Radiative processes are synchrotron radiation and IC scattering, 
and the cooling rate is denoted by $\dot{\gamma}_e$.
The escape time of electrons, $t_\mathrm{esc}$, is assumed to be $3 R/c$, 
where $c$ is the speed of light.
To calculate the spectrum of the primary electrons that emit radio to IR,
the injection spectrum of electrons is given by
\begin{equation}
  q_\mathrm{inj}(\gamma_e) = Q_e \gamma_e^{-\alpha_e}
  H(\gamma_e - \gamma_{e, \mathrm{min}}) H(\gamma_{e, \mathrm{max}} - \gamma_e) ,
\end{equation}
where $H(x)$ is the Heaviside function defined as $H(x) = 1$ for $x > 0$ and
$H(x) = 0$ for $x < 0$.
Here we do not specify the acceleration mechanisms to produce the injection spectrum.

In the same knot, nonthermal protons produce electrons
via collisions with soft photons.
The protons obey a power law given by
\begin{equation}
  n_p(\gamma_p) = 
    K_p \gamma_p^{-\alpha_p} H(\gamma_p-\gamma_{p, \mathrm{min}})
    H(\gamma_{p, \mathrm{max}}-\gamma_p) ,
\end{equation}
where $\gamma_p$ is the proton Lorentz factor and $n_p(\gamma_p)$ is the proton
number density per unit interval of $\gamma_p$.
We do not solve the kinetic equation for protons, 
because the cooling time of protons is much longer than that of electrons.
Then the production spectrum of electrons is calculated for a given proton spectrum
according to \citet{ka2008}.
To calculate the second population of electrons that emit the X-rays,
$q_\mathrm{inj}(\gamma_e)$ in Equation (\ref{eq:e-kinetic-hadronic}) is replaced 
by the electron/positron production spectrum by Bethe-Heitler process
and photo-pion processes based on \citet{ka2008}.
The soft photons used in electron production are supplied by radio-IR emission 
via above mentioned synchrotron radiation by the primary electrons.
In addition to these soft photons emitted in the knot, radiation from the AGN
and CMB is included.
The distance of the knot from the AGN core is denoted by $d_\mathrm{knot}$.
To calculate the soft photon injection rate from the AGN,
we use the photon spectrum compiled by NED.
Though the dispersion is large in the photon spectrum, 
we use a polynomial fit in $\log \nu$-$\log F_\nu$ plot of the NED spectrum.
As was mentioned in \S \ref{sec:intro},  
the beamed direction of AGN photons is not necessarily peaked 
to our line of sight and knots may receive a larger flux than that observed by us.
When we introduce the enhancement of the AGN photons due to this effect,
we use a parameter $f_A$ to denote the enhancement factor of 
the AGN photon number flux per unit frequency.  
We also consider the frequency shift of AGN photons by the Doppler effect $f_A^{1/2}$.
The effect of CMB photons on electron production is not important in our model, 
because the relativistic bulk motion of the knot is not considered.
The produced electrons emit photons by synchrotron radiation 
and IC scattering.
Here the latter contribution to the X-rays is negligible.
In this work proton synchrotron is also included.
However, because the magnetic field is weak, e.g.,  $\sim 0.1$ mG,
proton synchrotron has only minor effect on the emission spectrum.

Parameters for numerical calculations are  $R$, $B$, $d_\mathrm{knot}$, $f_A$, 
$K_p$, $\gamma_{p, \mathrm{min}}$, $\gamma_{p, \mathrm{max}}$, $\alpha_p$, 
$Q_e$, $\gamma_{e, \mathrm{min}}$, $\gamma_{e, \mathrm{max}}$, and $\alpha_e$.
We set $\gamma_{p, \mathrm{min}} = 1$ throughout the paper.

%%%%%%%%%%%%%%%%%%%%%%%%%%%%%%%%%%%%%%%%%%%%%%%%
\section{RESULTS}  \label{sec:results}

\subsection{3C 273}

%%%%%%%%%%%%%%%%%%%%%%%%
\begin{figure}[ht!]   % Figure 1
%  \epsscale{0.7}
%  \centering\includegraphics[scale=0.5,clip]{3c273-spec-5kpc-core-pion.eps}
  \centering\includegraphics[scale=0.5,clip]{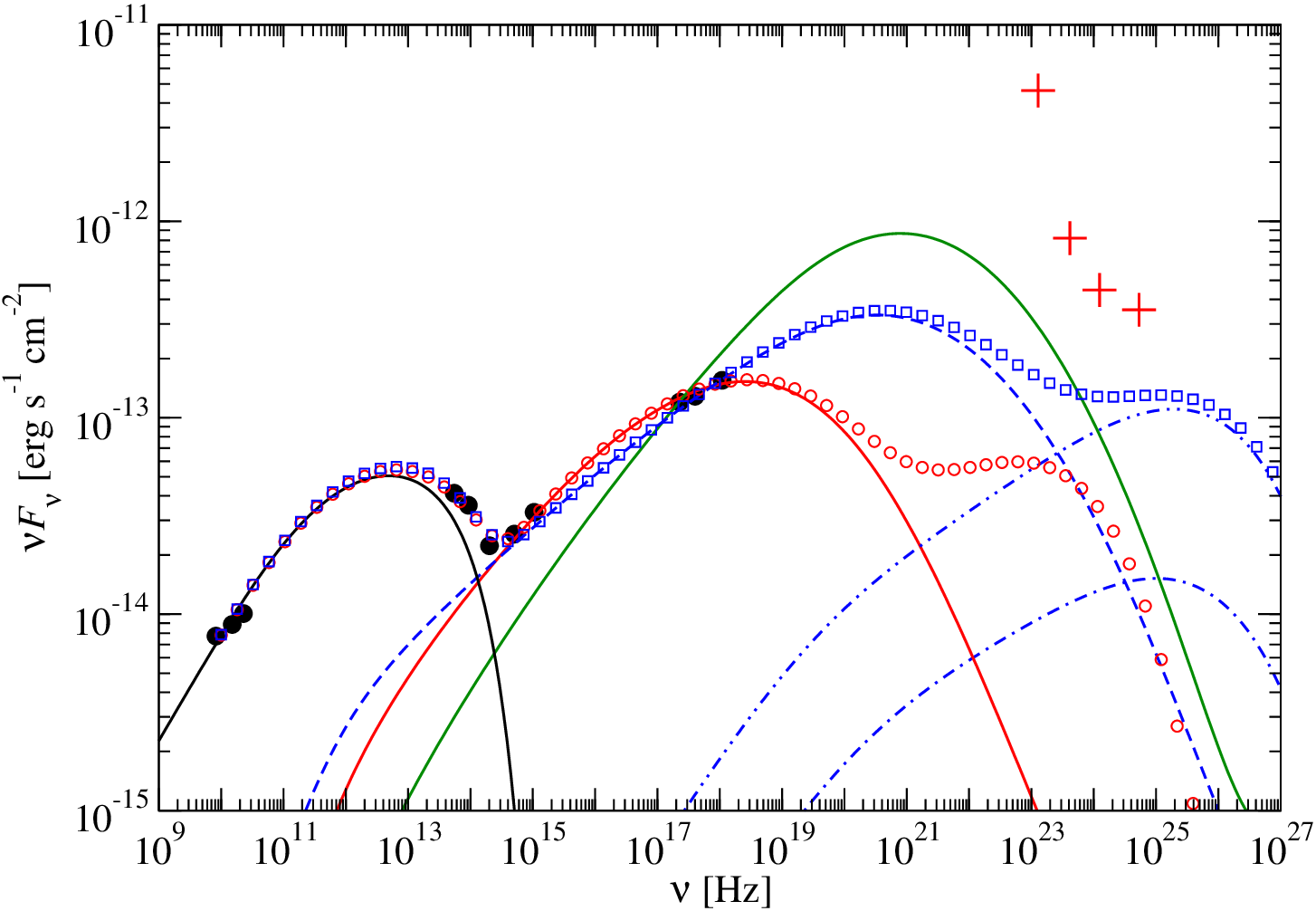}
  \caption{The spectral energy distribution of 3C 273 knots A and B1 combined
    (filled circles).
    Crosses are the upper limits set by \textit{Fermi}.
    The models are produced for $R=5$ kpc and $B=0.1$ mG.
    The solid black line is synchrotron radiation by primary electrons
    (parameters are given in Table \ref{tab:primary-el}).
    The solid green line is synchrotron radiation by electrons/positrons
    by Bethe-Heitler process for $\alpha_p=2$, 
    $\gamma_{p, \mathrm{max}}=10^{10}$, and $K_p=6 \times 10^{-4}$ cm$^{-3}$ 
    without AGN photons.
    The dashed blue line is for $\alpha_p =2.3$, $\gamma_{p, \mathrm{max}}=10^{10}$, 
    and $K_p = 0.154$ cm$^{-3}$ with AGN photons ($f_A=1$).
    For these parameters, synchrotron spectra by 
    positrons and electrons from photo-pion processes are shown 
    by dot-dot-dashed and dot-dashed blue lines, respectively.
    The total emission of these spectra is shown by blue open squares.
    Solid red line is for $\alpha_p=1.9$, $\gamma_{p, \mathrm{max}}=5 \times 10^8$,
    and $K_p = 1.77 \times 10^{-4}$ cm$^{-3}$ with AGN photons ($f_A=1$).
    The total emission for these parameters is shown by red open circles.
    The bump around $10^{23}$ Hz is the contribution from synchrotron radiation
    by positrons from photo-pion processes.
    \label{fig:3C273-spec-R5kpc}}
\end{figure}
%%%%%%%%%%%%%%%%%%%%%%%%

\begin{deluxetable*}{cccccc}[t!]
\tablecaption{Parameters for the primary electrons with $f_A = 1$ \label{tab:primary-el}}
\tablecolumns{6}
%\tablenum{1}
\tablewidth{0pt}
\tablehead{
\colhead{ } & \colhead{$R$} & \colhead{$B$} & \colhead{$\alpha_e$} & 
\colhead{$\gamma_{e, \mathrm{min}}$} & \colhead{$\gamma_{e, \mathrm{max}}$} \\
\colhead{ } & \colhead{(kpc)} & \colhead{(mG)} & \colhead{} & \colhead{} & \colhead{} 
}
\startdata
3C 273 knot A \& B1 & 5 & 0.1  & 1.9  & 10  &  $7 \times 10^5$ \\
3C 273 knot A \& B1 & 0.5 & 0.1  & 2.4  & 10  &  $9 \times 10^5$ \\
1136--135 knot A & 5 & 0.1  & 2.1 & 10 & $10^6$  \\
%1136--135 knot A & 5 & 0.05 & 2.2  & 10 & $10^6$  \\
1136--135 knot B & 5 & 0.1  & 1.9  & 10 & $9 \times 10^5$ \\
%1136--135 knot B & 5 & 0.05 & 2    & 10 & $9 \times 10^5$ \\
\enddata
%\tablecomments{The parameters for the primary electrons that emit radio-IR photons.}
\end{deluxetable*}

The jet of 3C 273 has been frequently observed at various wavelengths
\citep[see][for review]{mar2017}.
The $\gamma$-ray upper limits  on the knots A and B1 
observed  by \textit{Fermi} are shown in \citet{meyer2015}.
Though previous observations did not set strong constraints on the size of 
the emission regions,
\cite{mar2017} recently resolved the X-ray emission region down to $\sim 0.5$ kpc.
Considering that their result needs further confirmation,
in this work we  make models for $R = 5$ kpc and 0.5 kpc.

\subsubsection{Models with $R=5$ kpc}

We first show our models for $R = 5$ kpc and $B=0.1$ mG in Figure
\ref{fig:3C273-spec-R5kpc}.
The observed data are from \cite{meyer2015}.
The emission spectrum is fitted to the spectrum of  knots A and B1 combined.
The parameters for primary electrons that emit radio-IR are given in 
Table \ref{tab:primary-el}.
The injection spectrum is harder than the steady state electron spectrum 
because of radiative cooling.
The X-ray spectrum strongly depends on $\alpha_p$ and $\gamma_{p, \mathrm{max}}$.
However, because the data for the X-rays are limited, 
we cannot uniquely determine the values of these parameters.
In Figure \ref{fig:3C273-spec-R5kpc}, we show models with and without AGN photons.
The solid green line is for a model without AGN photons, where
$\alpha_p=2$, $\gamma_{p, \mathrm{max}}=10^{10}$, and $\gamma_{p, \mathrm{min}} = 1$.
(In this model photo-pion processes are not taken into account.)
When the photons from the AGN are not included, 
the synchrotron spectrum by electrons produced by Bethe-Heitler process is harder 
than the observed one at optical to X-rays.
We find that this is true as long as $\alpha_p \lesssim 2.7$ 
when $\gamma_{p, \mathrm{max}}=10^{10}$.
(Compared to PKS 0637--752, the observed spectrum of 3C 273 at optical to X-ray 
is softer.)
Then for  $\alpha_p \lesssim 2.7$, 
it is difficult to fit optical and X-rays simultaneously.
On the other hand, when the value of $\alpha_p$ is larger, 
energy requirement becomes larger.

\begin{figure}[ht!]   % Figure 2
%  \epsscale{0.7}
%  \centering\includegraphics[scale=0.5,clip]{radiation_field-3C273.eps}
  \centering\includegraphics[scale=0.5,clip]{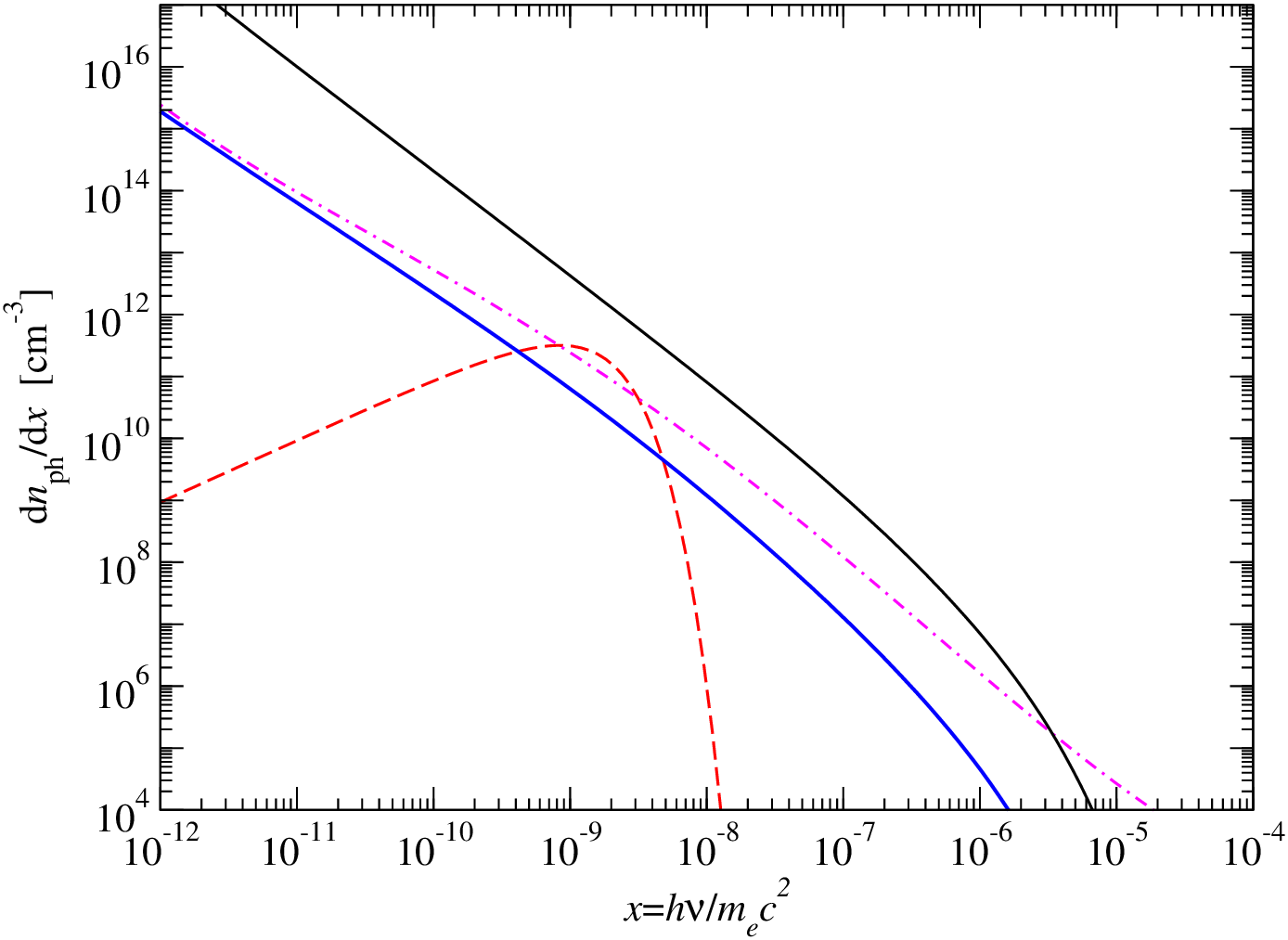}
  \caption{The target photon spectra in a knot of 3C 273.
    The black and blue solid lines are the radiation at radio-IR band 
    emitted by the primary electrons
    in knots A and B1 combined for $R=0.5$ kpc and $5$ kpc, respectively.
    The dashed line is the radiation field of CMB and the dot-dashed line is radiation
    emitted by the AGN core assuming $d_\mathrm{knot} = 50$ kpc.
    \label{fig:3C273-target-photon}}
\end{figure}

When the AGN photons are included, the value of $\alpha_p$ can be smaller.
Here we assume $d_\mathrm{knot} = 50$ kpc to determine the photon flux from the AGN.
Since the angle between the line of sight and the jet direction is not known,
the value of $d_\mathrm{knot}$ in our models 
should be taken as a characteristic value.
The spectra of target photons in electron production are shown 
in Figure \ref{fig:3C273-target-photon}.
The AGN photon spectrum is based on the data from NED.
For the most part we assume that knots receive the same flux of AGN photons
as the observed one, though we take into account for the distances of knots from the AGN.
In the following we assume $f_A=1$ and the effects of the beaming of AGN photons
are discussed in \S \ref{agn-beam}.
For $d_\mathrm{knot} = 50$ kpc, the photons from the AGN core dominate when $R = 5$ kpc.
We first fix the value of $\gamma_{p, \mathrm{max}} = 10^{10}$ to calculate 
the electron production rate.
We find that $\alpha_p$ must be smaller than 2.4 to fit the data.
When $\alpha_p \gtrsim 2.4$ with a fixed value of $\gamma_{p, \mathrm{max}} = 10^{10}$,
the electron spectrum becomes steeper and the number density of electrons 
with $\gamma_e \sim 10^5$ becomes larger.
As a result the flux at optical becomes too large.
When $\alpha_p = 2.3$, for example, 
$K_p = 0.15$ cm$^{-3}$ explains the observed spectrum
as depicted by the dashed blue line for Bethe-Heitler process
in Figure \ref{fig:3C273-spec-R5kpc}.
Radiation from positrons and electrons from photo-pion origin is shown by dot-dot-dashed 
and dot-dashed blue lines, respectively.
Their contribution is mainly above 10 GeV and does not exceed the Fermi limits.
The sum is shown by blue squares.
When the value of $\alpha_p$ is much smaller,
the value of $\gamma_{p, \mathrm{max}}$ should be smaller too to fit the data.
When $\alpha_p = 1.9$, we find $\gamma_{p, \mathrm{max}} = 5 \times 10^{8}$ with
$K_p = 1.8 \times 10^{-4}$ cm$^{-3}$ fit the data
as depicted by the solid red line for Bethe-Heitler process
in Figure \ref{fig:3C273-spec-R5kpc}.
The sum of Bethe-Heitler and photo-pion processes is shown by red squares.
This smaller value of $\gamma_{p, \mathrm{max}}$ is chosen to suppress 
over-production of X-rays. 
The $\gamma$-ray flux in our model is well below the \textit{Fermi} upper limits,
as is the case for PKS 0637--752 \citep{kt2017}.
When $\alpha_p \lesssim 1.8$, the spectrum at optical-IR is too hard
irrespective of the value of $\gamma_{p, \mathrm{max}}$.

%%%%%%%%%%%%%%%%%%%%%%%%
%%%%%%%%%%%%%%%%%%%%%%%%

\begin{figure}[ht!]   % Figure 3
%  \epsscale{0.7}
%  \centering\includegraphics[scale=0.5,clip]{el-spec-5kpc-core-pion2.3.eps}
  \centering\includegraphics[scale=0.5,clip]{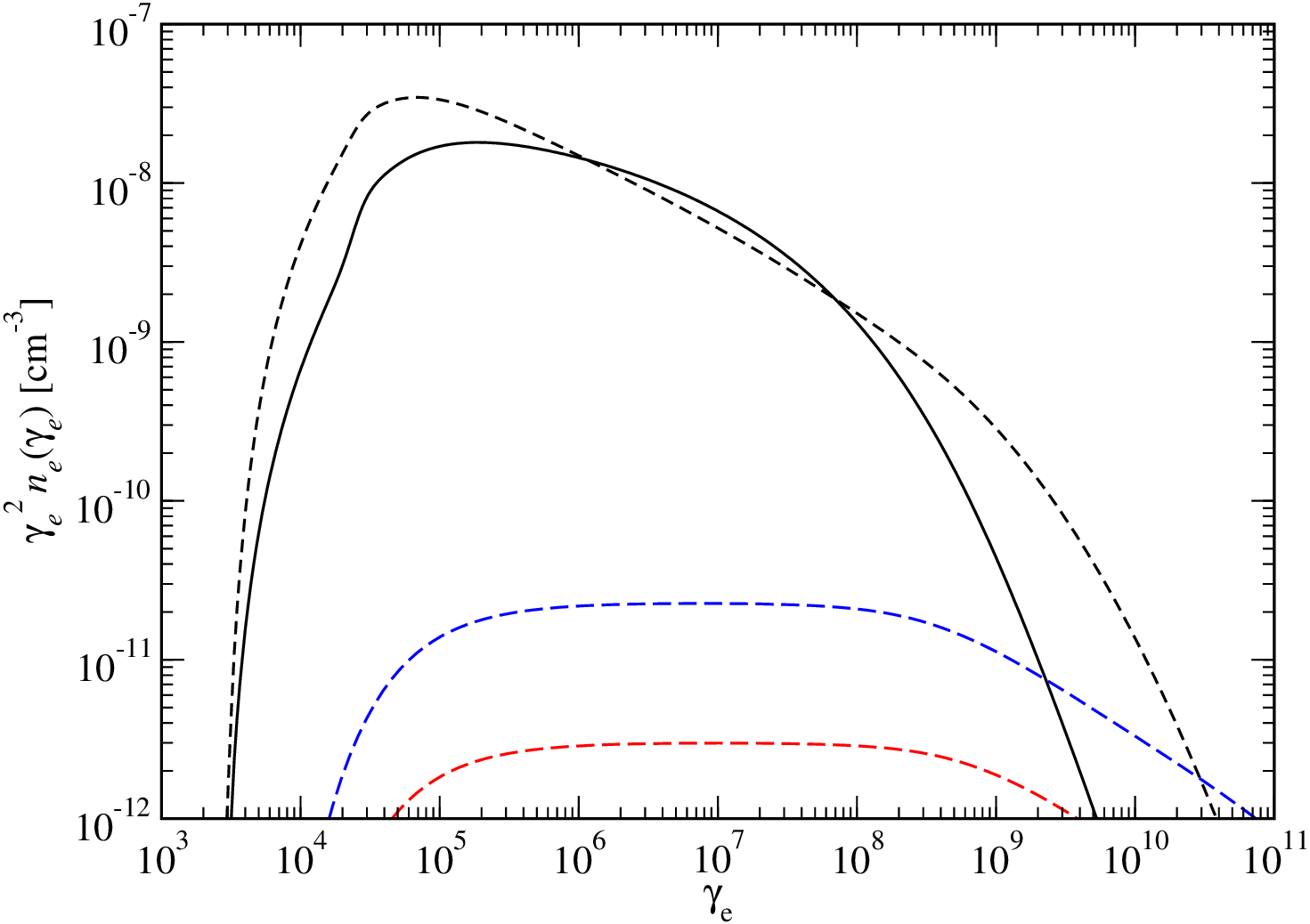}
  \caption{The steady state spectra of electrons produced by hadronic processes
    in a knot of 3C 273 with $f_A=1$.
    Electron/positron spectra produced by Bethe-Heitler process 
    are shown by the solid line 
    for $\alpha_p = 1.9$ and $\gamma_{p, \mathrm{max}} = 5 \times 10^8$
    and the dashed line for $\alpha_p = 2.3$ 
    and $\gamma_{p, \mathrm{max}} = 10^{10}$.
    The spectra of positrons and electrons via photo-pion processes are shown 
    by the blue and red dashed lines, respectively, for the model with $\alpha_p=2.3$ 
    and $\gamma_{p, \mathrm{max}} = 10^{10}$.
    \label{fig:3C273-electron-spec}}
\end{figure}

The electron spectra for the models with $R=5$ kpc and AGN photons
are shown in Figure \ref{fig:3C273-electron-spec}.
In the production spectrum of electrons, $\gamma_e^2 n_e(\gamma_e)$ 
from Bethe-Heitler process has a peak
at $\gamma_e \sim 2 \times 10^8$ for $\alpha_p =1.9$ and 
$\gamma_{p, \mathrm{max}} = 5 \times 10^8$.
Because of radiative cooling the peak is located at the lower energy,
$\gamma_e \sim 10^5$, in the steady state.
In the figure, the spectra of electrons and positrons by photo-pion processes are also shown.

%%%%%%%%%%%%%%%%%%%%%%%%
%%%%%%%%%%%%%%%%%%%%%%%%

%%%%%%%%%%%%%%%%%%%%%%%%
\begin{figure}[ht!]   % Figure 4
%  \epsscale{0.7}
%  \centering\includegraphics[scale=0.5,clip]{fit-F_nu-3C273-with-var.eps}
  \centering\includegraphics[scale=0.5,clip]{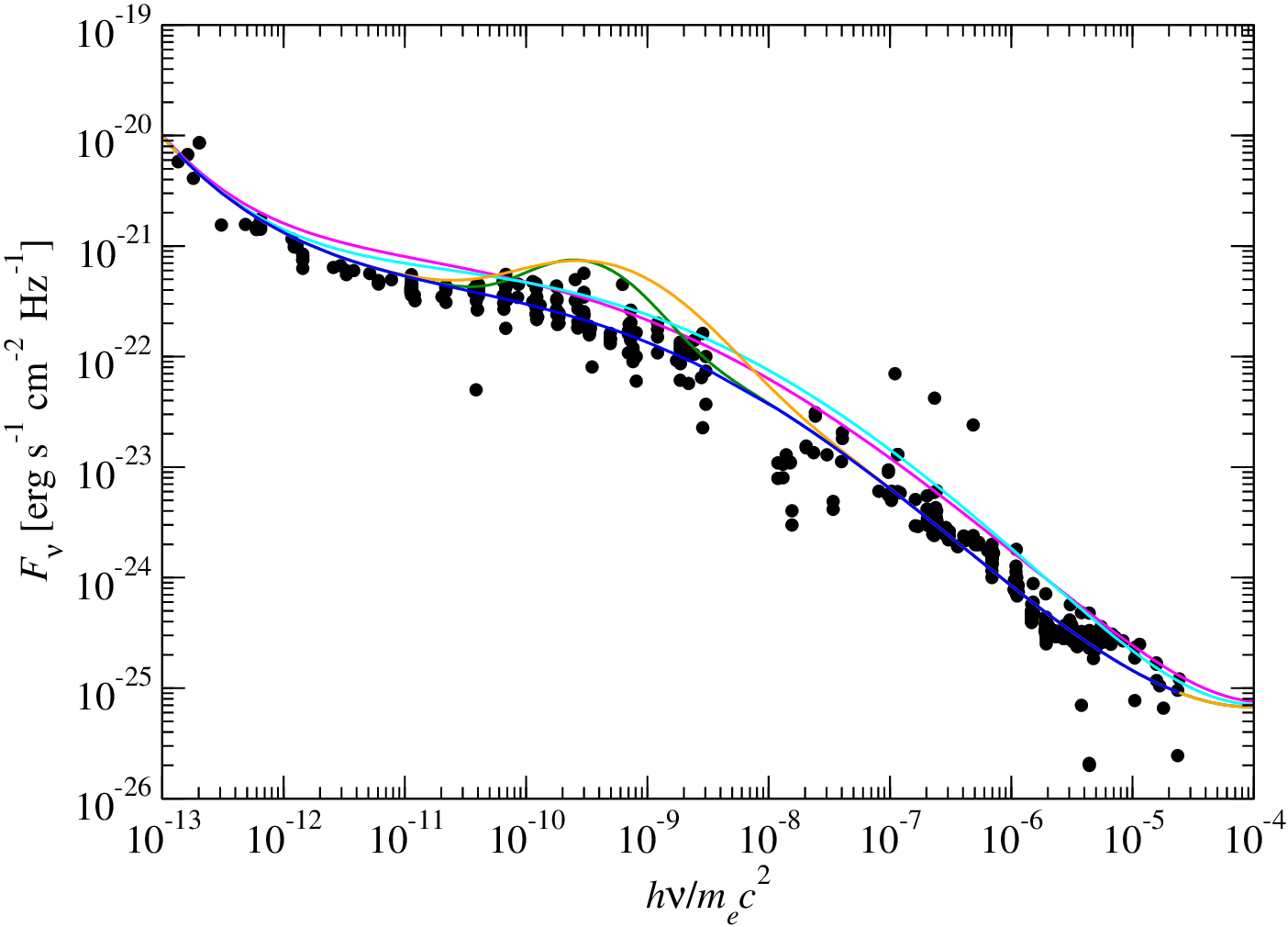}
  \caption{The radiation spectrum of 3C 273.
    Filled circles are the data from NED.
    The blue line is used to construct our models.
    Other various lines are used in Figure \ref{fig:3C273-SED-various-AGN}.
    \label{fig:3C273-various-AGN}
  }
\end{figure}
%%%%%%%%%%%%%%%%%%%%%%%%

%%%%%%%%%%%%%%%%%%%%%%%%
\begin{figure}[ht!]  % Figure 5
%  \epsscale{0.7}
%  \centering\includegraphics[scale=0.5,clip]{3c273-spec-5kpc-core-var.eps}
  \centering\includegraphics[scale=0.5,clip]{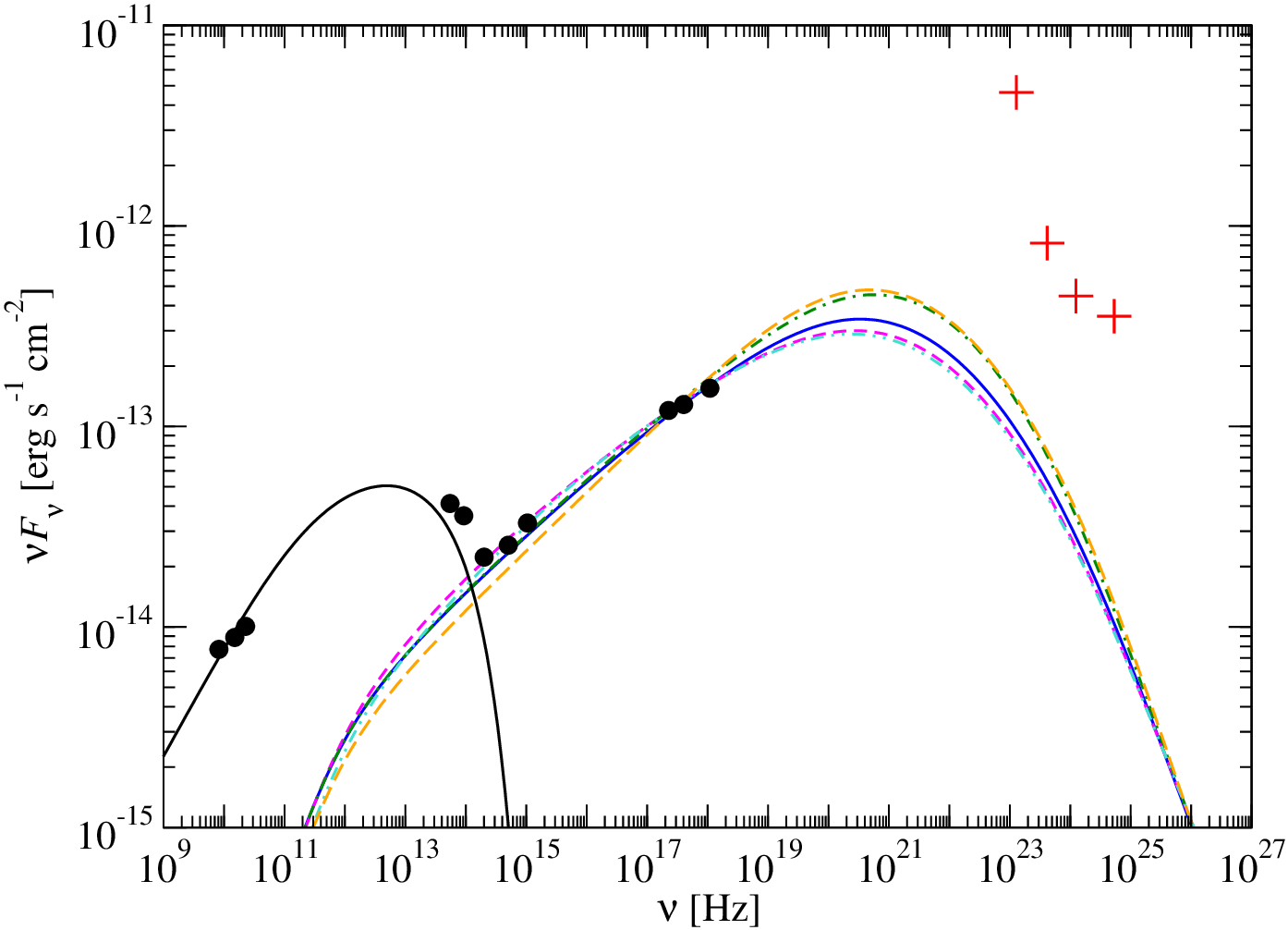}
  \caption{The spectral energy distribution of knots A and B1 combined of 3C 273 
    for various target photon spectra supplied by the AGN with $f_A=1$
    (Fig. \ref{fig:3C273-various-AGN}). 
    Here $\alpha_p = 2.3$ and $\gamma_{p, \mathrm{max}} =10^{10}$, 
    $R= 5$ kpc, $B=0.1$ mG, and $d_\mathrm{knot} = 50$ kpc. 
    The blue line is the same as shown in Figure \ref{fig:3C273-spec-R5kpc} 
    for $\alpha_p=2.3$.
    Other colors of lines correspond to the same colors shown 
    in Figure \ref{fig:3C273-various-AGN}.
    For simplicity, photo-pion processes are not shown. 
    \label{fig:3C273-SED-various-AGN}}
\end{figure}
%%%%%%%%%%%%%%%%%%%%%%%%

As mentioned above, we used the NED data for the AGN spectrum.
The NED data of the photon spectrum  have a large dispersion.
To examine the effect of the NED photon spectrum 
we artificially varied the photon spectrum
from the AGN (Figure \ref{fig:3C273-various-AGN}).
The results are shown in Figure \ref{fig:3C273-SED-various-AGN} 
for $\alpha_p = 2.3$ and $\gamma_{p, \mathrm{max}}=10^{10}$.
It is found that the X-ray spectrum does not strongly depend on the AGN spectrum.

%%%%%%%%%%%%%%%%%%%%%%%%
\begin{figure}[ht!]   % Figure 6
%  \epsscale{0.7}
%  \centering\includegraphics[scale=0.5,clip]{3c273-spec-05kpc-core.eps}
  \centering\includegraphics[scale=0.5,clip]{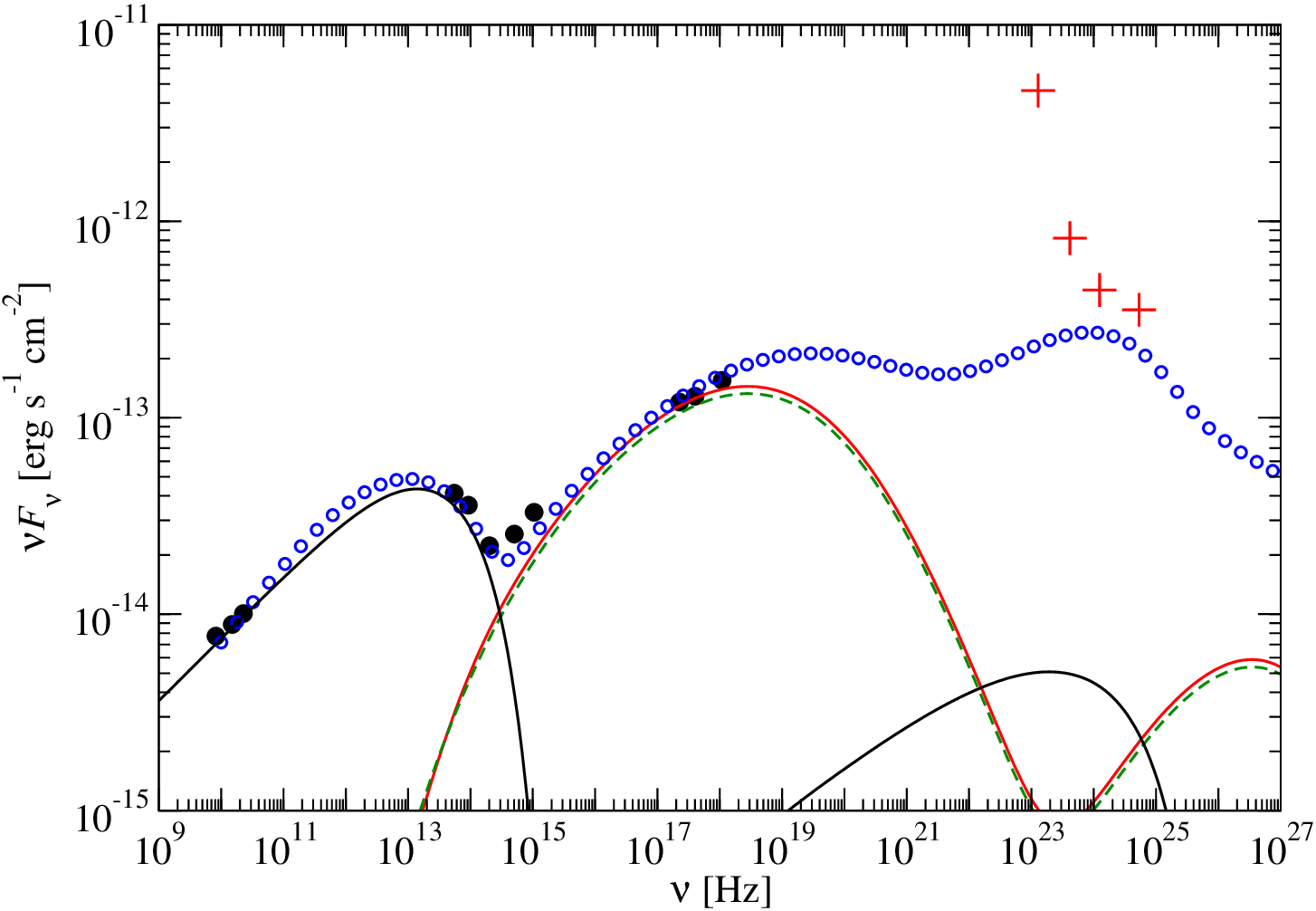}
  \caption{Models of 3C 273 for $R = 0.5$ kpc and $B=0.1$ mG.
    Black solid line: Synchrotron and synchrotron self-Compton emission 
    by the primary electrons.
    Radiation by electrons/positrons produced by Bethe-Heitler process is calculated
    for $\alpha_p = 1.9$ and $\gamma_{p, \mathrm{max}} = 5 \times 10^8$.
    Red solid line: $f_A=1$, $K_p = 1.68 \times 10^{-2}$ cm$^{-3}$,
    and $d_\mathrm{knot} = 50$ kpc.
    Green dashed line: Without AGN photons, and other parameters are the same as those 
    for the red solid line.
    With AGN photons and $f_A=40$, the sum of emission by
    the primary electrons and electrons/positrons by Bethe-Heitler and photo-pion 
    processes are shown by open circles,
    where $\alpha_p = 1.98$, $\gamma_{p, \mathrm{max}}=2 \times 10^9$, and 
    $K_p = 5.5 \times 10^{-4}$ cm$^{-3}$. 
    \label{fig:3C273-SED-R05kpc}}
\end{figure}
%%%%%%%%%%%%%%%%%%%%%%%%

\subsubsection{Models with $R=0.5$ kpc}

According to \citet{mar2017}, the size of knot A is resolved down to $\sim 0.5$ kpc
for the X-ray emission region.
We present models for $R = 0.5$ kpc and $B = 0.1$ mG in Figure \ref{fig:3C273-SED-R05kpc},
assuming that the spatial extent of radio-IR knots is 0.5 kpc as well.
When $R= 0.5$ kpc and $d_\mathrm{knot} = 50$ kpc,
photons from the AGN core are not important
as long as $f_A =1$ and CMB photons have only minor 
effects (Fig. \ref{fig:3C273-target-photon}).
The injection spectrum of the primary electrons is softer ($\alpha_e = 2.4$)
than that for the models with $R = 5$ kpc.
In Figure \ref{fig:3C273-SED-R05kpc}, we show a model with $\alpha_p = 1.9$,
$\gamma_{p, \mathrm{max}}= 5 \times 10^8$, and $K_p = 1.68 \times 10^{-2}$  cm$^{-3}$.
Compared to the model with $R=5$ kpc, the value of $K_p$ is increased by $10^{2}$.
As $R$ decreases the contribution of the radiation from the AGN decreases 
while the radio-IR photon density in a knot increases.
We also calculated a model with $\alpha_p = 2.3$ and 
$\gamma_{p, \mathrm{max}}= 10^{10}$ as for $R=5$ kpc and find that $K_p = 18$ cm$^{-3}$ fits
the observed X-rays.

\subsubsection{Proton power}

The proton energy, $U_p$, in a knot is calculated for our models with $f_A = 1$.
For $\alpha_p=2.3$, $K_p = 0.15$ cm$^{-3}$, 
$\gamma_{p, \mathrm{max}} = 10^{10}$, and $R = 5$ kpc, 
we obtain $U_p \sim 1.2 \times 10^{64}$ erg.
For $\alpha_p=1.9$, $K_p = 1.8 \times 10^{-4}$ cm$^{-3}$, 
$\gamma_{p, \mathrm{max}} = 5 \times 10^{8}$, and $R = 5$ kpc, 
the proton energy is $\sim 2.6 \times 10^{62}$ erg.
If the proton power is estimated as $L_p \sim U_p/(3 R/c)$,
where $3 R/c$ is the escape time (or adiabatic cooling time) of protons,
these models result in $L_p \gtrsim  10^{50} \, \text{erg} \, \text{s}^{-1}$.
The black hole mass $M_\mathrm{BH}$ of 3C 273 is estimated to be $\sim 10^8 M_\odot$
\citep[][and references therein]{esp2008},
while \citet{pt2005} estimate $M_\mathrm{BH} \sim 7 \times 10^{9} M_\sun$.
It is found that $L_p \sim 10^2$ - $ 10^4 L_\mathrm{Edd}$ 
for $M_\mathrm{BH} = 10^{10}$-$10^8 M_\sun$, 
$\alpha_p=1.9$, $K_p = 1.8 \times 10^{-4}$ cm$^{-3}$, 
and $\gamma_{p, \mathrm{max}} = 5 \times 10^{8}$.
Here $L_\mathrm{Edd} \sim 1.26 \times 10^{47} (M/10^9 M_\sun)$ erg s$^{-1}$ 
is the Eddington luminosity.
When $R = 0.5$ kpc, $U_p = 2.5 \times 10^{61}$ erg and
$L_p \sim 1.6 \times 10^{50}\,\text{erg} \, \text{s}^{-1} \sim 10^3 L_\mathrm{Edd}$
for $M_\mathrm{BH} = 10^9 M_\sun$, $\alpha_p = 1.9$, 
$\gamma_{p, \mathrm{max}} = 5\times 10^8$,  and $K_p = 1.7 \times 10^{-2}$ cm$^{-3}$.
$L_p$ does not much depend on the value of $R$ and
$K_p$ is nearly proportional to $R^{-2}$ when $\alpha_p$ is fixed.
The details of the energy density, $u_p$, and power of protons as well as the
energy density of the primary electrons, $u_e$, 
are shown in Table \ref{tab:proton-power}.
These models need the proton power larger than the Eddington luminosity by 
a few orders of magnitude and are hard to realize.

\subsubsection{Beaming of AGN photons} \label{agn-beam}

The required large proton power may be eased if the following situation is realized.
Though the AGN core of 3C 273 is known to exhibit super luminal motion,
the beam direction of radiation from the core 
can be slightly off the line of sight  and is towards the kpc-scale jets.
Then we observe a smaller flux of the AGN radiation
and the knots receive the  enhanced flux by the relativistic beaming effect.
In the following we assume that a knot receives the AGN photon number flux 
per unit frequency enhanced by $f_A$ compared with the flux 
that we receive and the frequency is shifted by $f_A^{1/2}$ in the knot frame.
To reduce the required large power of protons to as small as $10^{47}$ erg s$^{-1}$, 
the Doppler factor of the AGN core about a few times larger than we observe is necessary.
Note that the radiation spectrum of 3C 273 core is not like a typical blazar,
which is consistent with this hypothesis.
Here we remark that when there are enhanced AGN photons,
IC scattering off the AGN photons by the primary electrons is also enhanced.

\begin{figure}[ht!]   % Figure 7 New in revision
%  \epsscale{0.7}
%  \centering\includegraphics[scale=0.5,clip]{3c273-spec-5kpc-beam-f-various.eps}
  \centering\includegraphics[scale=0.5,clip]{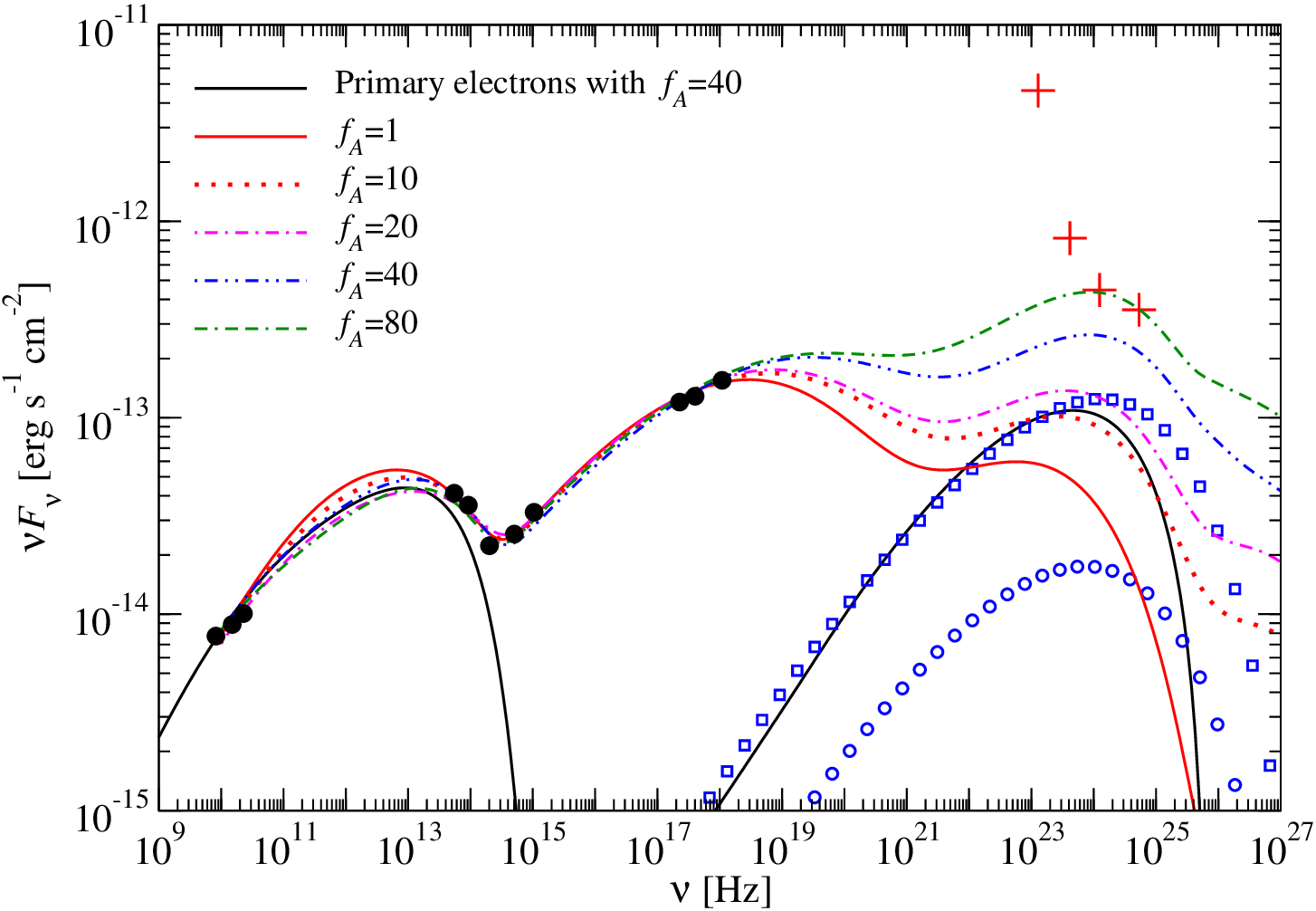}
  \caption{Models of 3C 273 for $R=5$ kpc and $B=0.1$ mG 
    with $f_A = 1$, 10, 20, 40, and 80.
    The spectra are the sum of  IC scattering off AGN photons by primary electrons
    and synchrotron radiation by primary electrons and 
    electrons/positrons produced by Bethe-Heitler and photo-pion processes.
    Note that for $f_A=1$, there is no contribution from IC scattering
    by primary electrons.
    The values of $\alpha_p$ and $\gamma_{p, \mathrm{max}}$ are given in
    Table \ref{tab:proton-power-agn100}.
    For $f_A=40$, emission by primary electrons is shown by a solid black line 
    and the spectra by positrons and electrons via photo-pion processes
    are shown by open squares and circles, respectively.
    There is a significant contribution of IC scattering 
    off AGN photons by primary electrons for $f_A \gtrsim 20$.
     \label{fig:3C273-spec-R5kpc-f-various}}
\end{figure}

Results for $f_A= 1$, 10, 20, 40, and 80 with $R=5$ kpc and  $B=0.1$ mG
are shown in Figure \ref{fig:3C273-spec-R5kpc-f-various}.
The values of $\alpha_p$ and $\gamma_{p, \mathrm{max}}$ are given in
Table \ref{tab:proton-power-agn100}.
The injection spectrum of the primary electrons are varied according to the value 
of $f_A$.
As the value of $f_A$ becomes larger, a harder injection spectrum is applied 
because radiative cooling becomes more effective by IC scattering off AGN photons.
The contribution of IC scattering off AGN photons by primary electrons
is shown in Figure \ref{fig:3C273-spec-R5kpc-f-various} for $f_A=40$.
The peak of IC scattering appears in the GeV region.
Synchrotron emission by electrons/positrons via photo-pion processes is also shown for 
$f_A = 40$.
For various values of $R$ and $f_A$, the energy density and power of protons are 
given in Table \ref{tab:proton-power-agn100}.
The upper limit of the value of $f_A$ is set by IC scattering by primary 
electrons and synchrotron radiation of positrons.
When $f_A \lesssim 80$, synchrotron emission by Bethe-Heitler electrons 
can explain the X-rays without violating the \textit{Fermi} limits.
The necessary proton power can be as small as $L_\mathrm{Edd}$ 
for large enough values of $f_A$. 
The visual luminosity of 3C 273 is $\sim 10^{46}$ erg s$^{-1}$ \citep{fan2009}.
If $f_A = 10$-40, the enhanced luminosity becomes $\sim 10^{48}$-$10^{49}$ erg s$^{-1}$
and is near the most luminous blazar  \citep[e.g.][]{ghi2017}.
When $R = 0.5$ kpc, the effect of AGN photons is weak if $f_A = 1$ 
as shown in Figure \ref{fig:3C273-target-photon}.
However, if $f_A \gtrsim 10$, the AGN photons are effective to reduce the proton power.
These effects of the AGN photons are shown in Figure \ref{fig:3C273-SED-R05kpc} for $f_A = 40$.

%%%%%%%%%%%%%%%%%%%%%%%%
%%%%%%%%%%%%%%%%%%%%%%%%

\subsubsection{Summary of 3C 273}

In this subsection we have presented a hadronic model to explain the X-ray emission
from the kpc-scale jet in 3C 273.
The X-rays are produced by synchrotron radiation emitted by electrons
produced by Bethe-Heitler process.
The target photons are radio-IR photons emitted by primary electrons in the kpc-scale
jet and CMB as well as the AGN photons.
The latter photons are important if $R = 5$ kpc 
but become less important if $R$ is smaller when $f_A=1$.
The required proton power exceeds the Eddington limit by a few orders of magnitude.
However, if the kpc-scale jet receives enhanced beamed radiation from  the AGN core,
i.e., if 3C 273 is a slightly misaligned blazar,
our models do not need the super Eddington power for protons.

\begin{deluxetable*}{cccccc}[t!]
\tablecaption{Proton energy density and power of 3C 273 for $f_A =1$ 
  \label{tab:proton-power}}
\tablecolumns{6}
%\tablenum{1}
\tablewidth{0pt}
\tablehead{
\colhead{$R$} & \colhead{$u_e$} & \colhead{$\alpha_p$} & \colhead{$\gamma_{p, \mathrm{max}}$} 
& \colhead{$u_p$} & \colhead{$L_p$} 
\\
\colhead{(kpc)} & \colhead{(erg cm$^{-3}$)} & \colhead{} & \colhead{} 
& \colhead{(erg cm$^{-3}$)} 
& \colhead{($10^{50}$ erg s$^{-1}$)} 
}
\startdata
0.5 & $4.6\times 10^{-8}$ & 1.9 & $5 \times 10^8$ & $1.6 \times 10^{-3}$   & 1.6 \\
1   & $4.9 \times 10^{-9}$ & 1.9 & $5 \times 10^8$ & $6.6 \times 10^{-4}$   &2.6 \\
1   & $4.9 \times 10^{-9}$ & 2.3 & $10^{10}$ & $3.3 \times 10^{-2}$   & $1.3 \times 10^2$ \\
5   & $1.3 \times 10^{-11}$ & 1.9 & $5 \times 10^8$ & $1.7 \times 10^{-5}$    & 1.7 \\
5   & $1.3 \times 10^{-11}$ & 2.3 & $10^{10}$ & $7.7 \times 10^{-4}$   & $7.7 \times 10$ \\
\enddata
\tablecomments{The magnetic energy density is 
  $u_\mathrm{mag} = 4.0 \times 10^{-10}$ erg cm$^{-3}$ for $B = 0.1$ mG.}
\end{deluxetable*}

\begin{deluxetable*}{cccccc}[t!]
  \tablecaption{Proton energy density and power of 3C 273
    for various values of $R$ and $f_A$
  \label{tab:proton-power-agn100}}
\tablecolumns{6}
%\tablenum{1}
\tablewidth{0pt}
\tablehead{
\colhead{$R$} & \colhead{$f_A$} & \colhead{$\alpha_p$} & \colhead{$\gamma_{p, \mathrm{max}}$} 
& \colhead{$u_p$} & \colhead{$L_p$} 
\\
\colhead{(kpc)} & \colhead{} & \colhead{} & \colhead{} & \colhead{(erg cm$^{-3}$)} 
& \colhead{(erg s$^{-1}$)} 
}
\startdata
0.5 & 10 & 1.98 & $9 \times 10^8$ & $3.1 \times 10^{-4}$   & $3.1 \times 10^{49}$ \\
0.5 & 20 & 1.95 & $10^9$ & $7.1 \times 10^{-5}$   & $7.1 \times 10^{48}$ \\
0.5 & 40 & 1.98 & $2 \times 10^9$ & $2.2 \times 10^{-5}$   & $2.2 \times 10^{48}$ \\
1   & 10 & 1.97 & $10^9$ & $3.6 \times 10^{-5}$  & $1.4 \times 10^{49}$  \\
1   & 20 & 1.95 & $10^{9}$ & $9.3 \times 10^{-6}$   & $3.7 \times 10^{48}$ \\
5   & 10 & 1.95 & $9 \times 10^8$ & $2.7 \times 10^{-7}$   & $2.7 \times 10^{48}$ \\
5   & 20 & 1.95 & $10^{9}$ & $7.5 \times 10^{-8}$   & $7.5 \times 10^{47}$ \\
5   & 40 & 1.97 & $2 \times 10^9$ & $2.0 \times 10^{-8}$   & $2.0 \times 10^{47}$ \\
5   & 80 & 1.97 & $2 \times 10^{9}$  & $5.9 \times 10^{-9}$   & $5.9 \times 10^{46}$ \\
\enddata
\end{deluxetable*}

\subsection{PKS 1136--135}

\begin{figure}[ht!]   % Figure 8 (old Figure 7)
%  \epsscale{0.7}
%  \centering\includegraphics[scale=0.5,clip]{1136-135-knotA-spec-d50kpc-cooling.eps}
  \centering\includegraphics[scale=0.5,clip]{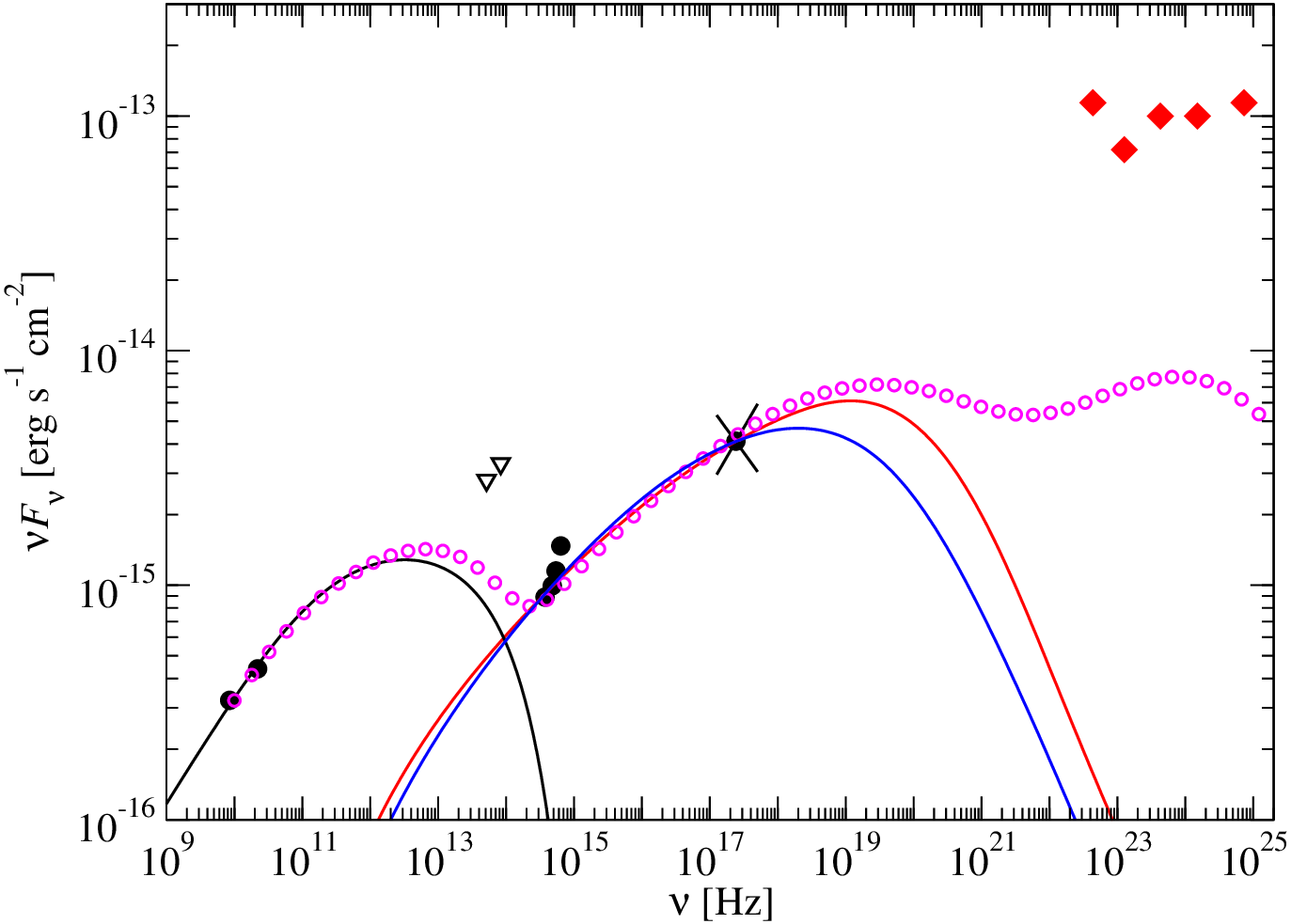}
  \caption{The spectral energy distribution 
    of knot A of PKS 1136--135 (filled circles).
    Triangles and diamonds are upper limits.
    For the models $R = 5$ kpc, $B = 0.1$ mG, and $d_\mathrm{knot} = 50$ kpc are assumed.
    Solid red line: $f_A=1$, $\alpha_p=2.1$,  $\gamma_{p, \mathrm{max}}=10^9$, 
    and $K_p = 3.1 \times 10^{-3}$ cm$^{-3}$.
    Solid blue line: $f_A=1$,  $\alpha_p=1.9$, 
    $\gamma_{p, \mathrm{max}}=5 \times 10^8$, and $K_p =10^{-4}$ cm$^{-3}$.
    The emission spectrum for $f_A = 40$ is shown by open circles,
    where $\alpha_p=1.98$, $\gamma_{p, \mathrm{max}}=2 \times 10^9$, 
    and $K_p = 3.7 \times 10^{-7}$ cm$^{-3}$.
    This spectrum is the sum of IC scattering  by primary electrons and 
    synchrotron radiation by primary electrons and electrons/positrons 
    produced by Bethe-Heitler and photo-pion processes.
    \label{fig:1136-knotA-R5kpc}}
\end{figure}

\begin{figure}[ht!]   % Figure 9  (old Figure 8)
%  \epsscale{0.7}
%  \centering\includegraphics[scale=0.5,clip]{1136-135-knotB-spec-cool.eps}
  \centering\includegraphics[scale=0.5,clip]{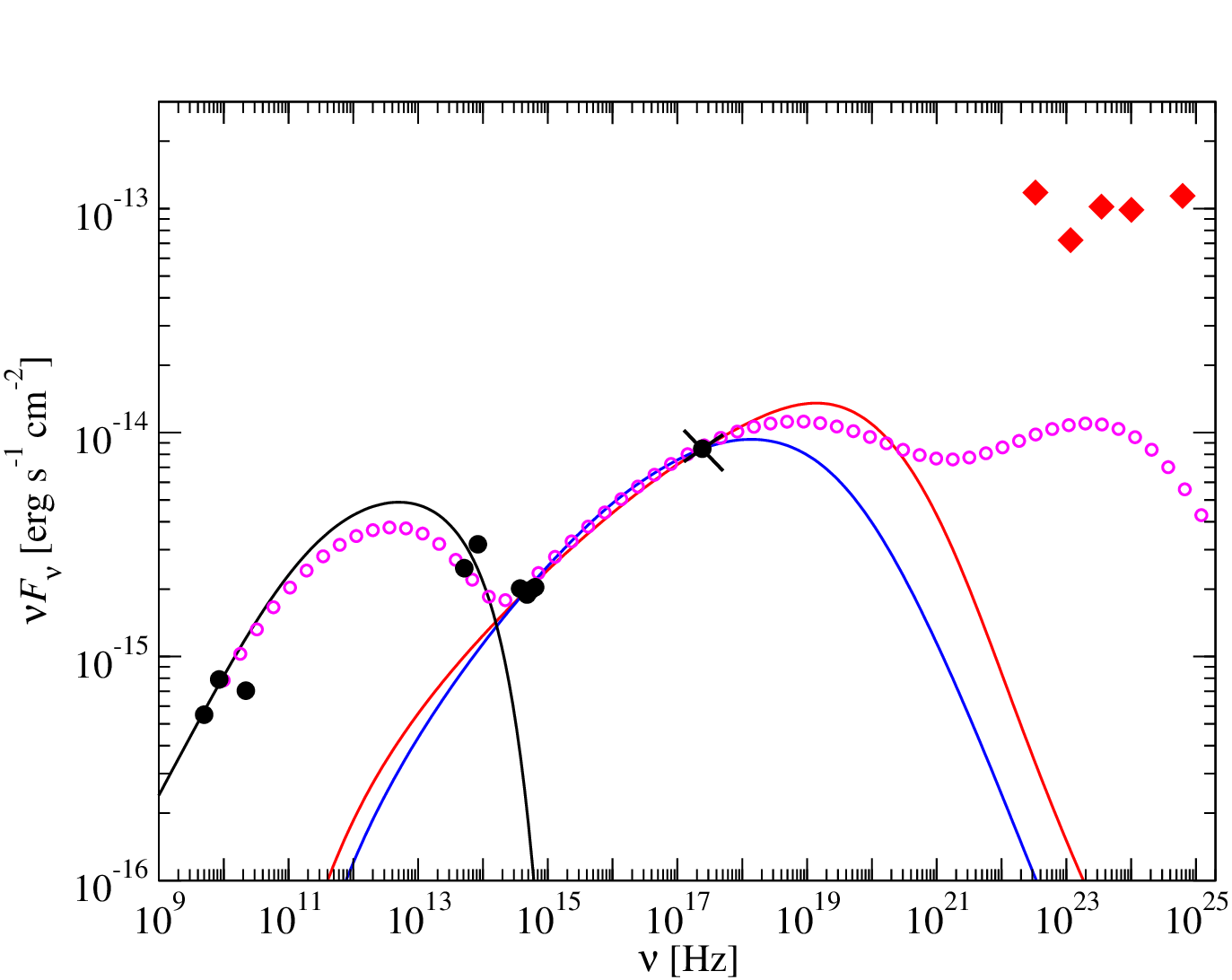}
  \caption{The spectral energy distribution of knot B of PKS 1136--135 (filled circles).
    The red diamonds are the \textit{Fermi} upper limits.
    For the models $R = 5$ kpc, $B = 0.1$ mG, and $d_\mathrm{knot} = 75$ kpc are assumed.
    Red solid line: $f_A=1$, $\alpha_p = 2.2$,  $\gamma_{p, \mathrm{max}} = 10^9$,
    and $K_p = 6.57 \times 10^{-2}$ cm$^{-3}$.
    Blue solid line: $f_A=1$, $\alpha_p = 1.9$,  $\gamma_{p, \mathrm{max}} = 4 \times 10^8$,
    and $K_p = 3.97 \times 10^{-4}$ cm$^{-3}$.
    The emission spectrum for $f_A = 40$ is shown by open circles,
    where $\alpha_p =1.98$, $\gamma_{p,\mathrm{max}}=10^9$,
    and $K_p = 2.1\times 10^{-6}$ cm$^{-3}$.
    This spectrum is the sum of IC scattering by primary electrons 
    and synchrotron radiation by primary electrons and electrons/positrons 
    produced by Bethe-Heitler and photo-pion processes.
    \label{fig:1136-knotB-R5kpc}}
\end{figure}

\begin{figure}[ht!]   % Figure 10   (old Figure 9)
%  \epsscale{0.7}
%  \centering\includegraphics[scale=0.5,clip]{fit-F_nu-1136-135.eps}
  \centering\includegraphics[scale=0.5,clip]{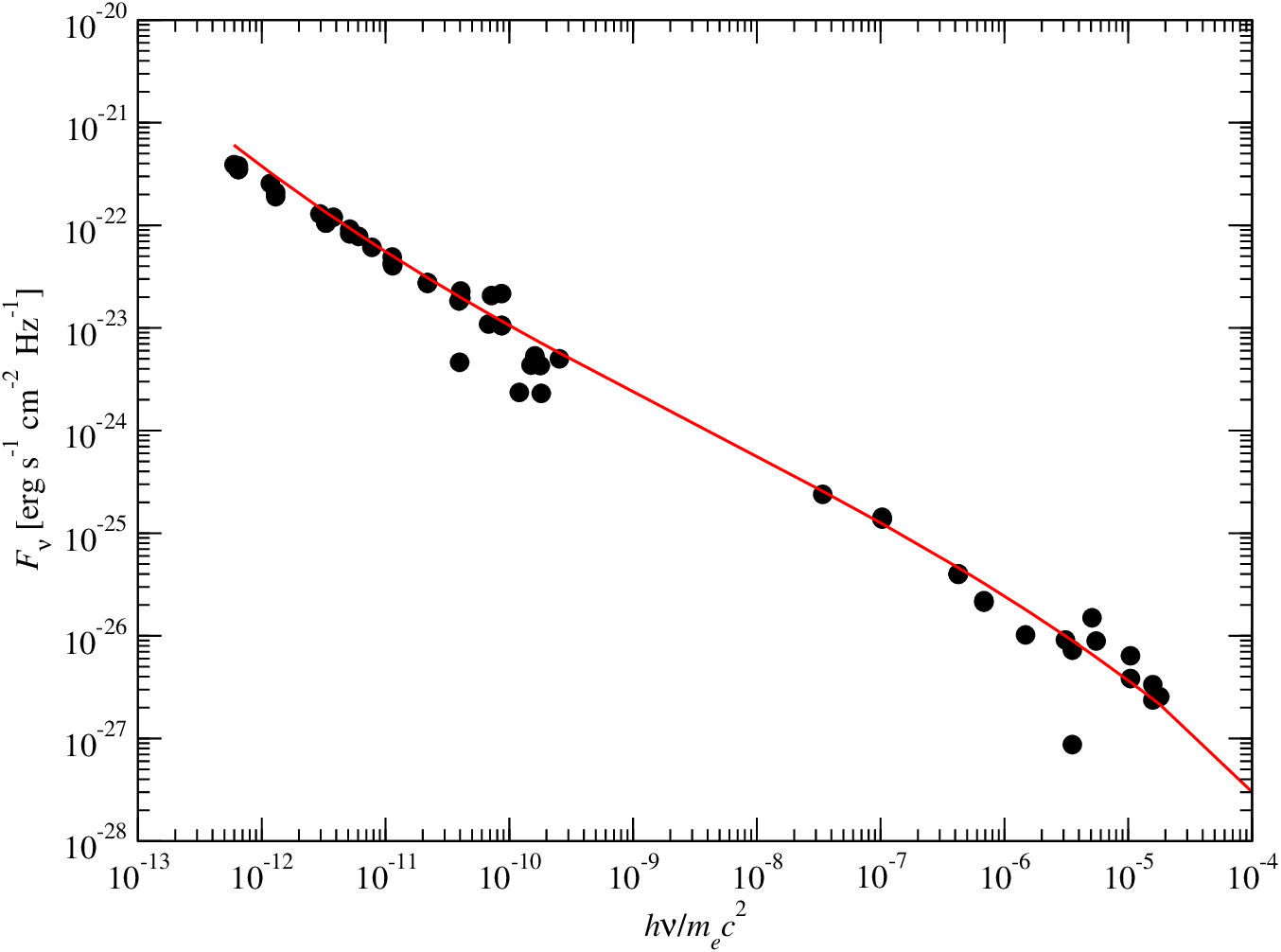}
  \caption{The radiation spectrum of PKS 1136--135.
    Filled circles are the data from NED.
    The red line is a polynomial fit of the $\log \nu$-$\log F_\nu$ representation
    and used in calculations of electron/positron production.
    \label{fig:1136-135-AGN}}
\end{figure}

\begin{figure}[ht!]   % Figure 11   (old Figure 10)
%  \epsscale{0.7}
%  \centering\includegraphics[scale=0.5,clip]{radiation_field-1136-135-paper.eps}
  \centering\includegraphics[scale=0.5,clip]{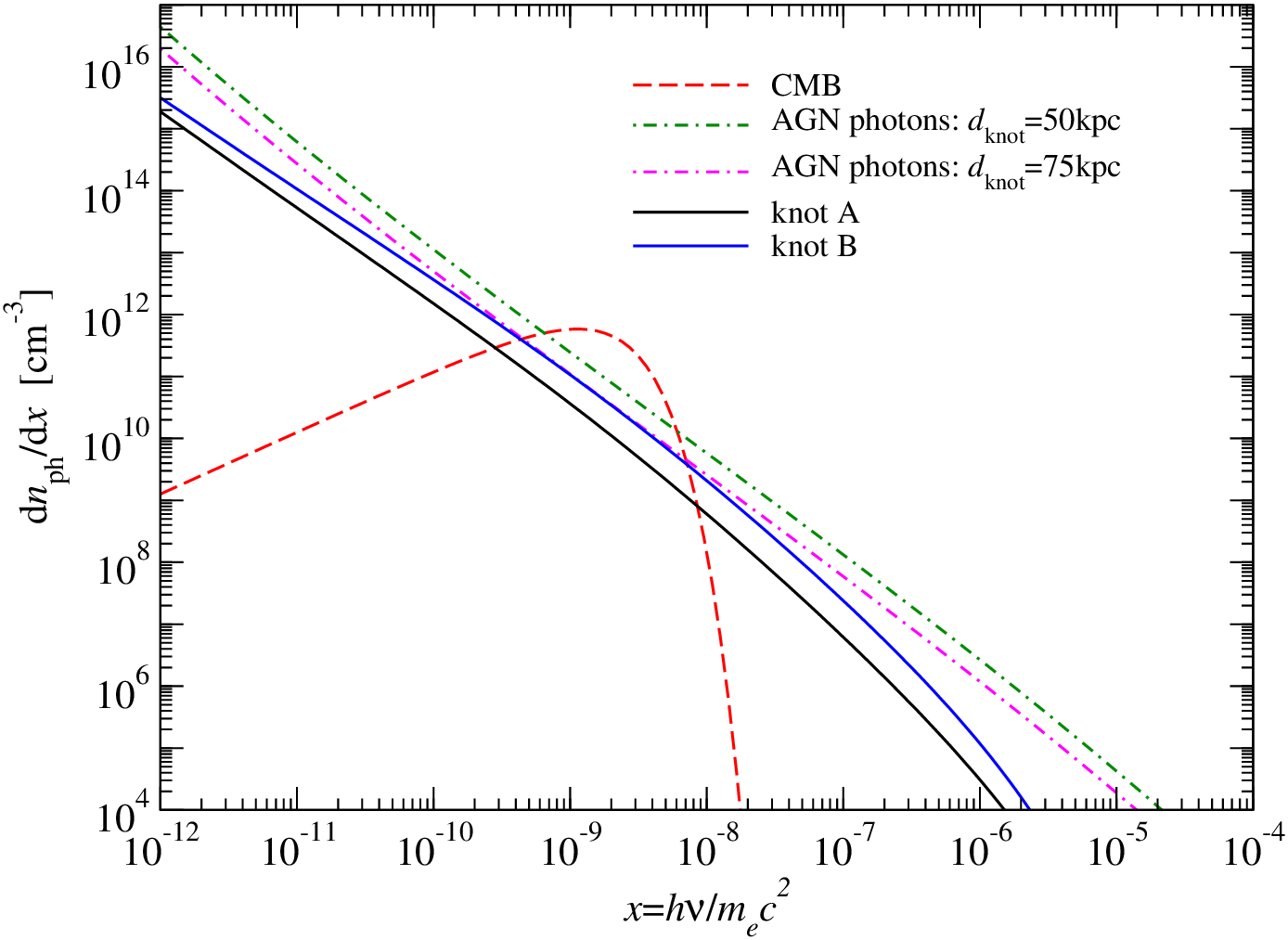}
  \caption{The target photon spectra for proton-photon collisions for PKS 1136--135.
    The synchrotron radiation by primary electrons in knot A and knot B are shown
    by the black and blue solid lines, respectively.
    CMB photons are shown by the red dashed line.
    The green and magenta dot-dashed lines are the AGN photons with $f_A=1$ 
    for $d_\mathrm{knot} = 50$ and 75 kpc, respectively.
    \label{fig:1136-135-target}}
\end{figure}

Another kpc-scale jet bright in X-rays is PKS 1136--135.
The observed spectra of PKS 1136--135 are shown in Figures \ref{fig:1136-knotA-R5kpc}
and \ref{fig:1136-knotB-R5kpc} for knots A and B, respectively.
The data are taken from \citet{cara2013}, \citet{sam2006}, and \citet{uchi2007}.
The \textit{Fermi} upper limits from \citet{bm2017} are shown by diamonds.
The AGN core has visual and NIR luminosities of 
$1.1 \times 10^{46} \, \text{erg} \, \text{s}^{-1}$ and 
$2.2 \times 10^{45} \, \text{erg} \, \text{s}^{-1}$, respectively \citep{uchi2007}.
The AGN spectrum is given in Figure \ref{fig:1136-135-AGN},
which is used to calculate the electron production rate in hadronic processes.
We assume $d_\mathrm{knot} = 50$ kpc and 75 kpc for knots A and B, respectively.
The contribution of soft photons from various sources is shown in
Figure \ref{fig:1136-135-target} for $f_A=1$.
In knot A, CMB photons dominate in $\nu \sim 8 \times 10^{10}$-$7 \times 10^{11}$ Hz 
and AGN photons dominate otherwise.
In knot B, CMB photons dominate in $\nu \sim 5 \times 10^{10}$-$10^{12}$ Hz
and AGN photons dominate otherwise. 
The parameters for primary electrons are given in Table \ref{tab:primary-el}.
For simplicity we fix $R = 5$ kpc in all models.
First, we describe knot A.
When $B = 0.1$ mG, we did not find a value of $\alpha_p$ to fit the data for
$\gamma_{p, \mathrm{max}}=10^{10}$,
because the flux at $\sim 10^{14}$ - $\sim 10^{15}$ Hz becomes too large.
When the value of $\gamma_{p, \mathrm{max}}$ is decreased to $10^9$, we find that
$\alpha_p = 2.1$ and $K_p = 3.1 \times 10^{-3}$ cm$^{-3}$ fit the X-ray data,
though still the slope at $\sim 10^{14}$ - $\sim 10^{15}$ Hz is too small compared to
the observed data.
When $\alpha_p = 1.9$ is assumed, $\gamma_{p, \mathrm{max}}= 5 \times 10^8$
and  $K_p = 10^{-4}$ cm$^{-3}$ fit the X-ray data.
%When $B = 0.05$ mG and $\gamma_{p, \mathrm{max}}=10^9$ are assumed,
%we obtain $\alpha_p = 2.1$ and $K_p = 2.73 \times 10^{-3}$ cm$^{-3}$.
Second, we show  the results for knot B.
Models with parameter sets
($\alpha_p = 2.2$, $\gamma_{p, \mathrm{max}}= 10^9$, $K_p = 6.57 \times 10^{-2}$ cm$^{-3}$) 
and ($\alpha_p = 1.9$, $\gamma_{p, \mathrm{max}}= 4 \times 10^8$, 
$K_p = 3.97 \times 10^{-4}$ cm$^{-3}$), for $B = 0.1$ mG, are shown in
Figure \ref{fig:1136-knotB-R5kpc}.
Since these models are not much different from knot A, 
we discuss only knot A below.

%%%%%%%%%%%%%%%%%%%%%%%%
%%%%%%%%%%%%%%%%%%%%%%%%

%%%%%%%%%%%%%%%%%%%%%%%%
%%%%%%%%%%%%%%%%%%%%%%%%

%%%%%%%%%%%%%%%%%%%%%%%%
%%%%%%%%%%%%%%%%%%%%%%%%

%%%%%%%%%%%%%%%%%%%%%%%%
%%%%%%%%%%%%%%%%%%%%%%%%

The observed spectrum of knot A is harder at $\nu \sim 5 \times 10^{14}$ Hz than 
our model spectra.
\citet{cara2013} fit the data with the electron spectral index 2.0 
and the large value of the minimum Lorentz factor of electrons, 
i.e., $\sim 3 \times 10^6$.
Unless some acceleration mechanisms exist, it may be difficult to keep this large value of
$\gamma_{e, \mathrm{min}}$ against rapid radiative cooling.
Otherwise, different emission regions might contribute.

For knot A with $f_A=1$, 
we obtain the proton energy $U_p \sim 1.5 \times 10^{62}$ erg
for $B=0.1$ mG, $\alpha_p = 1.9$, $K_p = 10^{-4}$ cm$^{-3}$, 
and $\gamma_{p, \mathrm{max}} = 5 \times 10^8$.
The proton power $L_p$ is $\sim 9.6 \times 10^{49}$ erg s$^{-1}$.
The black hole mass of PKS 1136--135 is estimated to be $4.6 \times 10^8 M_\odot$
\citep{osh2002}, so that $L_\mathrm{Edd} \sim 10^{47}$ erg s$^{-1}$
and $L_p \sim 10^3 L_\mathrm{Edd}$.
This value is similar to the case for 3C 273 
and the beaming effect of AGN core emission on the kpc knots is to be considered 
to suppress the super Eddington of the proton power.
When $f_A=40$ for knot A, we obtain $K_p = 3.7 \times 10^{-7}$ and 
$L_p \sim 1.5 \times 10^{47} \text{erg} \, \text{s}^{-1} \sim L_\mathrm{Edd}$.
The emission spectra of knots A and B for $f_A = 40$ are shown in 
Figures \ref{fig:1136-knotA-R5kpc} and  \ref{fig:1136-knotB-R5kpc} by circles.

\section{Summary and Discussion}  \label{sec:conclusion}

We examined the possibility to explain the X-rays from kpc-scale jets of  
3C 273 and PKS 1136--135 by synchrotron radiation emitted by electrons/positrons 
produced by Bethe-Heitler process.
We explored various conditions such as the size of the emission region,
the soft photon injection from the AGN core, the spectral shape of protons,
and the proton power.

Because the efficiency of electron/positron production by Bethe-Heitler process is low,
the proton power larger than the Eddington limit  is required 
to explain the X-rays from the knots of 3C 273 and PKS 1136--135
as for PKS 0637--752.
Observations show that there indeed exist AGNs with super-Eddington luminosity
\citep[e.g.,][and references therein]{jdn2017}.
However, our models for 3C 273 and PKS 1136--135 require the  proton power 
much larger than the Eddington limit, e.g., $\sim 10^2$-$10^4 L_\mathrm{Edd}$.
Then we presented a model with a reduced proton power, which assumes
that the radiation from the AGN core to the knots is enhanced by the relativistic beaming.
Here the axis of the AGN core jet is slightly off the line of sight,
so that the observed AGN flux is smaller than the flux received by the knots.
If the number flux per unit frequency received by the knots 
is about 10-40 times larger than
that observed by us, the required proton power becomes about the Eddington limit.
When this is the case, IC scattering by primary electrons also increases the flux in GeV band.
Since the sum of this contribution and emission by electrons/positrons from photo-pion processes
should be lower than the \textit{Fermi} upper limits, 
the upper bound of the enhancement factor $f_A$ is less than 
$\sim 80$ for 3C 273. For PKS 1136--135, $f_A$ can be larger because 
the constraint by the \textit{Fermi} limit is weaker.
According to \citet{sam2001} the brightness distribution between X-rays and radio
are different in 3C 273  \citep[see also][]{mar2017}.
While radio emission is brighter at the outer edge of the jet, 
X-rays are stronger at inner portion. This may imply the photons from the AGN core
play a role in X-ray emission in the large scale jet by Bethe-Heitler process.

As for the size of the emission region, in the hadronic model with enhanced AGN photons,
a smaller size model needs more proton power than a larger one for a given value 
of $f_A$.
For example, the value of $L_p$ of a knot with $R = 0.5$ kpc of 3C 273
is larger by a factor $\sim  10$ than that with $R = 5$ kpc.

We briefly comment on the effects of AGN photons on PKS0637--752, 
which was not covered in our previous paper.
As for PKS 0637--752, \citet{ed2006} reported superluminal motion of the pc-scale jet.
Core optical luminosity is about three orders of magnitude higher than that of the knot. 
The projected distance to the knot from the core is about 70 kpc 
so that the radiation from the core can be ignored if the beaming effect is neglected. 
Only if the core is not aligned with the line of sight and the kpc-scale jet is aligned
with beamed radiation, the core emission may be effective.
When we adopt a small viewing angle, the distance of the knot becomes as large as Mpc, 
which is unlikely. 
The spectral energy distribution of the core seems to be a little different from 
a typical blazar; 
the luminosity of the high energy component is mild \citep{meyer2017}.
Thus the possibility of a slightly off beamed jet cannot be neglected.  
If this is the case, the kpc-scale jet may see an order of magnitude larger beamed radiation.

The cascade of $\gamma$-rays produced by $\pi^0$ decay is not included in this work.
The optical depth of the electron-positron pair production in photon-photon collisions
is $\tau_{\gamma \gamma} \sim \sigma_\mathrm{T} n_\mathrm{soft} R$
for the threshold photon energies,
where $\sigma_\mathrm{T}$ is the Thomson cross section and $n_\mathrm{soft}$ 
is the target soft photon number density.
For a knot of 3C 273 with $R = 5$ kpc, $B=0.1$ mG, and $f_A = 40$, 
we find $\tau_{\gamma \gamma} \gtrsim 1$ for $\nu \gtrsim 3 \times 10^{24}$ Hz or
$h \nu \gtrsim 12$ GeV.
The produced $\gamma$-ray spectrum has a peak at $\nu\sim 10^{29}$ Hz 
and its production rate is similar to that of positrons.
Then electron-positron pairs produced by $\gamma$-rays emit synchrotron radiation
with a spectrum similar to that of positrons via photo-pion processes.
Thus the $\gamma$-ray flux may be as twice as that of positron synchrotron radiation.

In this work we did not consider the relativistic bulk motion of the knots towards us.
When the knots have the beaming factor larger than unity,
the soft photon density by primary electrons is reduced to account for the observed
radio-IR spectrum.  
This results in the lower electron production rate by Bethe-Heitler process.
Photons from the AGN core received by the knots also decrease,
while CMB radiation is enhanced in the knot frame and IC/CMB process becomes effective.

Our model does not satisfy the equipartition condition, i.e.,
the energy density of protons is much larger than that of magnetic fields.
Furthermore, we assumed the acceleration of protons is more efficient than that of
the primary electrons.
That is, the proton power is larger than the electron power, and the maximum Lorentz factor
of protons is also larger than that of the primary electrons.
However, these issues, equipartition and acceleration, are beyond the scope of this paper.

%%%%%%%%%%%%%%%%%%%%%%%%%%%%%%%%%%%%%%%%%%%%%%%%%%%%%%%%%%
\acknowledgments
This research has made use of the NASA/IPAC Extragalactic Database (NED) 
which is operated by the Jet Propulsion Laboratory, California Institute of Technology, 
under contract with the National Aeronautics and Space Administration.

%%%%%%%%%%%%%%%%%%%%%%%%%%%%%%%%%%%%%%%%%%%%%%%%%%%%%%%%%

%% This command is needed to show the entire author+affilation list when
%% the collaboration and author truncation commands are used.  It has to
%% go at the end of the manuscript.
%\allauthors

%% Include this line if you are using the \added, \replaced, \deleted
%% commands to see a summary list of all changes at the end of the article.
%\listofchanges

\end{document}